\documentclass[prl,twocolumn]{revtex4}
\usepackage{graphicx}
\usepackage{amsmath}
\usepackage{bbold}
\usepackage{amssymb}
\usepackage{color}
\usepackage[normalem]{ulem}
\usepackage{epsfig}
\usepackage{amssymb,amsfonts,amsmath,color}
\usepackage{array}
\usepackage{makecell}

\begin{document}

\title{Population dynamics in stochastic environments}

\author{Jayant Pande and Nadav M. Shnerb}

\affiliation{Department of Physics, Bar-Ilan University,
Ramat-Gan IL52900, Israel}

\begin{abstract}
\noindent Populations are made up of an integer number of individuals and are subject to stochastic birth-death processes whose rates may vary in time. Useful quantities, like the chance of ultimate fixation, satisfy an appropriate difference (master) equation, but closed-form solutions of these equations are rare. Analytical insights in fields like population genetics, ecology and evolution rely, almost exclusively, on an uncontrolled application of the diffusion approximation (DA) which assumes the smoothness of the relevant quantities over the set of integers. Here we combine asymptotic matching techniques with a first-order (controlling-factor) WKB method to obtain a theory whose range of applicability is much wider. This allows us to rederive DA from a more general theory, to identify its limitations, and to suggest alternative analytical solutions and scalable numerical techniques when it fails. We carry out our analysis for the calculation of the fixation probability in a fluctuating environment, highlighting the difference between (on average) deleterious and beneficial mutant invasion and the intricate distinction between weak and strong selection.
\end{abstract}

\maketitle

The dynamics of all biological populations take place in fluctuating environments. Micro-environmental variations affect individuals in an uncorrelated manner and are usually modeled by a stochastic birth-death process with fixed rates (usually termed demographic stochasticity or genetic drift). Macro variations may affect the fitness of entire types or strains, causing the rates themselves to vary in time (this being known as environmental stochasticity or fluctuating selection). Traditionally, the theory of population genetics and evolution was focused on the interplay between selection  and drift~\cite{parsons2010some}, assuming fixed birth and death rates. The effects of fluctuating selection were considered only rarely~\cite{takahata1975effect,takahata1979genetic}, despite the fact that they are known to be one of the main drivers of ecological dynamics~\cite{lande2003stochastic}. Recent empirical studies have documented periodic and stochastic coherent variations in relative fitness~\cite{bergland2014genomic,bell2010fluctuating,messer2016can} as well as variations in the birth and death rates~\cite{caceres1997temporal,hoekstra2001strength,leigh2007neutral,hekstra2012contingency,kalyuzhny2014niche,kalyuzhny2014temporal,chisholm2014temporal}. These findings have triggered a renewed interest in the effect of macro-environmental variations and varying selection coefficients in ecology, population genetics and evolutionary dynamics~\cite{mustonen2008molecular,huerta2008population,ashcroft2014fixation,cvijovic2015fate,hidalgo2017species,wienand2017evolution,danino2018fixation,meyer2018noise,wienand2018eco, marrec2020resist,shoemaker2020integrating}.

In any population the number of individuals is discrete, so the time evolution of the system is a stochastic process over the integers. This leads naturally to a difference (master) equation. Analytical solutions for such equations are rare, and numerical solutions are limited to relatively small systems and are case-specific. To obtain general insights, or address large systems, one has to rely on approximations that facilitate an analytical solution.

For many years, the (nearly) only game in town has been the diffusion approximation (DA)~\cite{crow1970introduction,karlin1981second}. Recent studies, including~\cite{danino2018fixation,meyer2018noise}, present a detailed DA-based analysis of species competition in fluctuating environments. However (as we shall discuss further below) the parameter range in which DA is applicable is quite limited. During the last few decades many authors have considered its limitations and employed alternate methods for a few problems in population dynamics~\cite{elgart2004rare,assaf2006spectral,uecker2011fixation,wienand2017evolution,marrec2020resist}. In particular, a WKB (large-deviations) technique, first used in population dynamics by Kessler and Shnerb~\cite{kessler2007extinction}, has become quite popular~\cite{ovaskainen2010stochastic,assaf2017wkb}. Existing attempts to use the WKB approach for generic systems with fluctuating environments are nevertheless limited, as they lead to an inherently two-dimensional problem that requires numerical solutions~\cite{assaf2017wkb}.

To overcome these limitations, we here combine the asymptotic matching method used in \cite{danino2018fixation,meyer2018noise} with the WKB technique and derive a theory that is general and whose range of applicability is much wider than that of DA. Armed with this method we present analytical expressions that work quite well when DA fails, quantify and clarify the different behaviors in weak-selection and strong-selection regimes, and suggest a new, scalable numerical approach. Our technique  may be applied to all types of models to calculate a wide range of interesting quantities like the  absorption time and the chance of invasion. Here we clarify its details and demonstrate its power for the calculation of the most important quantity in the theory of population genetics and evolution, the chance of ultimate fixation.

We consider an individual-based model with standard Wright-Fisher (non-overlapping generation) dynamics. The model describes the zero-sum competition of a mutant type (represented by $n$ individuals, of frequency $x=n/N$, in a haplotype population of size $N$) and a wild type (with $N-n$ individuals). In every generation (a year, say) all the individuals die and the chance of the mutant type to win each of the $N$ slots in the next generation is given by
\begin{equation}
r = \frac{xe^s}{1-x+xe^s},
\end{equation}
where $s$ is the selection parameter (log-fitness). In a fixed environment $s$ is time-independent. To model a fluctuating environment we allow $s$ to jump between two states (dichotomous noise), $s=s_0 \pm \sigma$, where the sign of $\sigma$ is picked randomly each generation.

For a system with many states indexed by $k$ (in our case $+\sigma$ corresponds to $k=1$ and $-\sigma$ to $k=2$), the chance of ultimate fixation when the mutant type has $n$ individuals and the environmental state is $k$ satisfies the discrete Backward Kolmogorov Equation (BKE),
\begin{equation}  \label{eqn1}
\Pi_n^k = \sum_{m,k'} W_{n,k \to n+m,k'} \Pi_{n+m}^{k'}.
\end{equation}
Here $W_{n,k \to n+m,k'}$ are the transition probabilities between states. Since $\sum_{m,k'} W_{n,k \to n+m,k'} =1$, any constant is a solution of (\ref{eqn1}), and the general solution is a linear combination of this constant and the nontrivial solution $\overline{ \Pi}$,  $\Pi_n^k = C_1 + C_2 \overline{ \Pi}_n^k$. $C_1$ and $C_2$ are determined by the boundary conditions $\Pi_0^k = 0$ and $\Pi_N^k = 1$. A direct numerical solution of (\ref{eqn1}) requires matrix inversion~\cite{danino2018fixation,comp}, so the numerical effort needed grows as $N^3$.

The diffusion approximation (DA) approach, as employed in old and recent studies~\cite{crow1970introduction,ewens2012mathematical,takahata1975effect,takahata1979genetic,danino2018fixation,meyer2018noise,meyer2020evolutionary}, relies on the smoothness of $\Pi$ over its state space, which, for a system with fluctuating selection, means both the population states $n$ and the environmental states $k$ (e.g., $\Pi_n^{\sigma} - \Pi_n^{-\sigma} \ll 1$~\cite{danino2018stability,danino2018fixation}). The latter requires the persistence time of the environment to be much smaller than the fixation time~\cite{mustonen2008molecular,cvijovic2015fate}. In what follows we assume that $N$ is large and this condition is satisfied, so we have to deal only with the ($N$-independent~\cite{kessler2007extinction}) non-smoothness of $\Pi$ over the population states. With this assumption one can average $\Pi_n^k$ over all the $k$ states to yield $\Pi_n$.

Under the DA assumption of a smooth-enough $\Pi$, one may approximate $\sum_{m} W_{n \to n+m} \Pi_{n+m}$ by
\begin{equation} \label{eqn1a}
\Pi(x=n/N) + \mathbb{E}(m)\Pi'(x)/N + \mathrm{Var}(m)\Pi''(x)/2N^2,
\end{equation}
where the quantities  $\mathbb{E}(m)/N$ and $\mathrm{Var}(m)/N^2$ are the (per-generation) mean and variance of $\Delta x$ for a given $n$, as calculated from the transition probabilities. A rigorous derivation of (\ref{eqn1a}), which clarifies why the variance (and not the second moment) appears in the last term, is provided in the companion paper~\cite{comp}.

DA fails when the gradient of $\Pi$ over the different states is too steep~\cite{kessler2007extinction}. In contrast, the first-order (controlling-factor) WKB approximation relies on a weaker condition: it requires  $\ln \Pi$ to be smooth over the integers~\cite{kessler2007extinction}. To use this method, one writes $\Pi_n = \exp(S_n)$ and uses $S_{n+m} \approx S_m + m S'(n)$. Since Eq.~(\ref{eqn1}) (with the transition probabilities $W$ averaged over all the different environmental states $k$) is linear, the $\exp(S_n)$ factor cancels out, giving
\begin{equation}  \label{eqn2}
\sum_{m} W_{n \to n+m} e^{q m} = 1,
\end{equation}
where $q \equiv S'(n)$. Once $q$ is calculated from Eq.~(\ref{eqn2}), the controlling factor  $S_n = \int q(n) dn$ is obtained by direct integration and yields the special solution $\overline{\Pi}_n = \exp(S_n)$.

Eq.~(\ref{eqn2}) is still a complicated transcendental equation and its numerical solution for every $n$ requires an effort comparable with a direct matrix inversion that solves Eq.~(\ref{eqn1}). To proceed we employ a \emph{two-destination approximation}. For any state of the system we calculate the expected state displacement in the next generation, $s_\text{e}$, and its variance, $\sigma_\text{e}^2$. Once these quantities are known,  Eq.~(\ref{eqn2}) is replaced by the corresponding WKB equation for a walk with only two, equally-probable destinations. This yields the  \emph{fundamental transcendental equation},
\begin{equation} \label{eqn2a}
\frac{1}{2}\left( e^{q(s_\text{e}+\sigma_\text{e})}+ e^{q(s_\text{e}-\sigma_\text{e})}\right) = e^{qs_\text{e}} \cosh(q \sigma_\text{e}) = 1.
\end{equation}

Eq.~(\ref{eqn2a}), too, has no solution in terms of elementary functions, but now we can suggest a powerful approximation scheme. For a given $x$ we define $\tilde{q} \equiv q(x) \sigma_\text{e}$ and $\tilde{s} \equiv s_\text{e}/\sigma_\text{e}$, so the fundamental equation takes the form
\begin{equation} \label{eqn4}
e^{\tilde{q} \tilde{s}} \cosh(\tilde{q}) = 1.
\end{equation}
Since $\tilde{q}(-\tilde{s}) = -\tilde{q}(\tilde{s})$, one need consider only one of the signs.

\begin{figure}[h]
\begin{center}
\includegraphics[width=6cm]{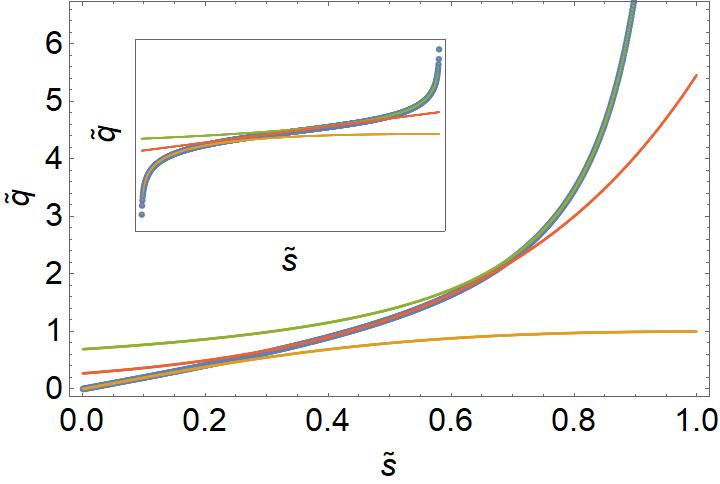}
\end{center}
\vspace{-0.5 cm}
\caption{Numerical solution of Eq.~(\ref{eqn4}) as a function of $\tilde{s}$ (blue), together with the low-$\tilde{q}$ approximation  (orange), the medium-$\tilde{q}$ approximation (red) and the large-$\tilde{q}$ approximation (green), as detailed in Table \ref{table1}. The agreement appears good on both the real (main) and semi-logarithmic (inset) scales.       \label{fig1m}}
\end{figure}

\begin{table}[h]
\caption{}
\centering
\begin{tabular}{|c | c |c|}
\hline\hline
 small $ \tilde{q}$ &  $|\tilde{s}| < 0.25$ & $q \approx -\frac{2 s_\text{e}}{(s_\text{e}^2+\sigma_\text{e}^2)}$  \\ \hline
medium $ \tilde{q}$ & $0.25< |\tilde{s}| < 0.7$ & $q \approx -{\rm sign}(s_\text{e}) \frac{e^{3 (s_\text{e}/\sigma_\text{e})-1.3}}{\sigma_\text{e}}$ \\ \hline
 large $ \tilde{q}$ & $0.7<|\tilde{s}|$   &  $q \approx \frac{\ln 2}{s_\text{e} - {\rm sign}(s_\text{e})\sigma_\text{e}}$  \\ \hline \hline
\end{tabular}
\label{table1}
\end{table}

The range $0 \le |\tilde{s}|<1$ is divided into three sectors.  When $\tilde{q}$ is small, one can expand (\ref{eqn4}) to second order in $\tilde{q}$ to obtain $\tilde{q} = -2\tilde{s}/(\tilde{s}^2+1)$. This approximation holds for $|\tilde{s}| < 0.25$. When $\tilde{q}$ is large the cosh function is replaced by $\exp(|\tilde{q}|)/2$ and $\tilde{q} \approx \ln 2/[\tilde{s} - \rm{sign}(\tilde{s})]$. This approximation works well when $|\tilde{s}| > 0.7$. In the medium-$q$ sector, $0.25<|\tilde{s}| < 0.7$, a good approximation is $\tilde{q} = -\rm{sign}(\tilde{s}) \exp(3 \tilde{s}-1.3)$. A comparison between these approximations and the direct numerical solution of Eq.~(\ref{eqn4}) is presented in Figure~\ref{fig1m}. The solutions, translated back to $q = \tilde{q}/\sigma_\text{e}$, are summarized in Table~\ref{table1}.

Now we determine $s_\text{e}$ and $\sigma_\text{e}$. In a pure drift (neutral) system, the mean change in $x$ per generation is zero, and the variance is given by $x(1-x)/N$. Accordingly, for a \emph{given} value of $s$ we approximate the $x$-dynamics by the sum of its deterministic change plus or minus the standard deviation associated with the drift,
\begin{eqnarray}
x &\to& \frac{xe^s}{1-x+xe^s} \pm \sqrt{\frac{x(1-x)}{N}},  \nonumber \\
1-x &\to& \frac{1-x}{1-x+xe^s} \mp \sqrt{\frac{x(1-x)}{N}}.
\end{eqnarray}

For convenience we now switch to the logit state variable $z \equiv \ln[x/(1-x)]$ (so $x = e^z/(1+e^z)$). The dynamics along the $z$-axis are relatively simple,
\begin{equation} \label{eqn5c}
z  \to z' \approx z+s \pm B(s), \text{ with } B(s) \equiv  \frac{1+\cosh(s+z)}{\sqrt{N} \cosh(z/2)}.
\end{equation}
Since $s$ takes the values $s_0 + \sigma$ and $s_0 - \sigma$, the calculation of the mean and the variance of $\Delta z$ (the change in $z$) involves four processes, each with probability $1/4$. The mean change in $z$ in one generation, $s_\text{e}$, equals simply $s_0$. The effective stochasticity $\sigma_\text{e} \equiv \sqrt{\mathrm{Var}(\Delta z)}$ is given by
\begin{equation} \label{eqn5b}
\sigma_\text{e}^2 = \frac{1}{4} \sum_{\substack{\zeta_1 = -1,1 \\ \zeta_2 = -1,1}} \left[ s_0 + \zeta_1 \sigma + \zeta_2 B(s_0 + \zeta_1 \sigma) \right]^2 - s_0^2.
\end{equation}

\begin{figure}[h]
\centering
\includegraphics[width=6cm]{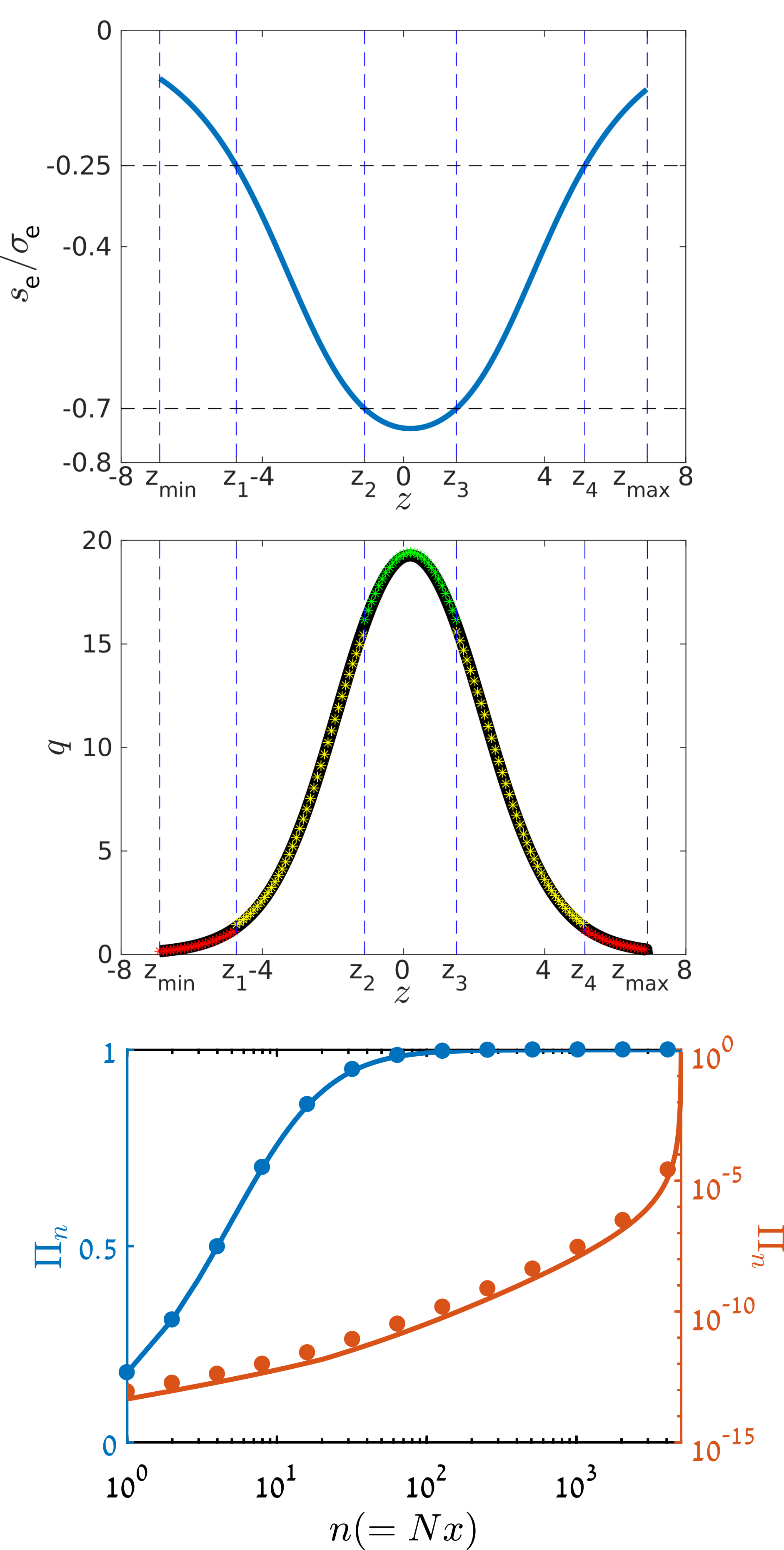}
\caption{\label{fig:scalable}  $s_\text{e}/\sigma_\text{e}$ (top) and $q$ (middle) plotted against $z \equiv \ln[x/(1-x)]$, between $z_\text{min} \approx -\ln N$ (one individual) to $z_\text{max} \approx \ln N$ ($N-1$ individuals),  for $s_0 = -0.1$, $\sigma = 0.12$ and $N = 1000$.  In the top panel, $z_1$ and $z_4$ are defined as the $z$-values for which $|s_\text{e}/\sigma_\text{e}| = 0.25$, and $z_2$ and $z_3$ are the $z$-values for which $|s_\text{e}/\sigma_\text{e}| = 0.7$. In the middle panel, the solid black line shows $q$ as extracted from the numerical solution of the fundamental equation (\ref{eqn4}), while the asterisks show the approximate analytical expressions in Table \ref{table1} (red: small $q$; yellow: medium $q$; green: large $q$). The bottom panel shows $\Pi_n$ as a function of $n$, as obtained from the direct numerical solution of the BKE (\ref{eqn1}) (circles) and from the scalable numerical solution (solid lines) whose details are given in the companion paper \cite{comp}. The blue curves (left $y$-axis, linear scale) are for $s_0 = 0.1$, and the orange curves (right $y$-axis, logarithmic scale) for $s_0 = -0.1$, with $\sigma = 0.3$ in each case.  }\label{fig2m}
\end{figure}

Once $s_\text{e}(z)$ and $\sigma_\text{e}(z)$ are known, even through complicated expressions, one can solve numerically for $\Pi$ with $N$-independent numerical effort, as explained in detail in the companion paper~\cite{comp}. This requires finding (at most) four regional boundaries by solving $\tilde{s} = 0.25$ and $\tilde{s} = 0.7$ for $z$, between $z_\text{min}$ and $z_\text{max}$ (see Figure~\ref{fig2m}). In each region $q$ is extracted from $\sigma_\text{e}$ and $s_\text{e}$ via Table~\ref{table1}, and then $\Pi(z)$ is calculated using (at most) five numerical integrations, one in each segment.

To use the asymptotic matching technique, $N$ must be sufficiently large, such that in the bulk the drift is negligible with respect to the environmental stochasticity. A WKB solution is then obtained separately for the inner ($x \ll 1$), outer ($1-x \ll 1$) and intermediate regimes, and the solutions are then matched using the same procedure that was employed for DA in \cite{danino2018fixation,meyer2018noise}.

As explained in the companion paper~\cite{comp}, an intermediate regime exists when $ \sqrt{N} \gg 2/(\sigma-|s_0|)$. In this regime Eq.~(\ref{eqn2a}) is solved with $s_\text{e} = s_0$ and $\sigma_\text{e} = \sigma$. The solution, $q_\text{mid}$, is  independent of $z$, so $\Pi$ in the intermediate regime takes the form
\begin{equation} \label{eqn7a}
\Pi_\text{mid}(x) = C_3 + C_2 \left( \frac{x}{1-x} \right)^{q_\text{mid}}.
\end{equation}

In the  inner regime, $x \ll 1$ so $z \approx \ln x$ is negative and large. In this limit $\cosh(z/2) \approx 1/(2\sqrt{x})$ and $\cosh(s + z) \approx \exp(-s)/(2x)$. Accordingly [see Eq.~(\ref{eqn5c})],
\begin{equation}
B(s) \approx \frac{2x+e^{-s}}{\sqrt{Nx}} \approx  \frac{e^{-s}}{\sqrt{n}}.
\end{equation}
Plugging this into Eq.~(\ref{eqn5b}) we get
\begin{equation}
\sigma_\text{e}^2 (n) = \sigma^2 + \frac{K(s_0)}{n}, \text{ with } K(s_0) \equiv e^{-2s_0}\cosh(2 \sigma).
\end{equation}

Once $q(z)$ [or $q(n)$] is determined using Table~\ref{table1}, the solution in the inner regime is given by
\begin{equation}
\Pi_\text{in} = C_1 \left(1-e^{\int_{-\infty}^z q(z') dz'}\right) = C_1 \left(1-e^{\int_{0}^n \frac{q(n')}{n'} dn'}\right).
\end{equation}

By symmetry, the outer regime satisfies
\begin{equation} \label{outer1}
\Pi(x|_{s_0}) = 1-\Pi(1-x|_{-s_0}),
\end{equation}
so
\begin{equation}
\Pi_\text{out} = 1-C_4 \left(1-e^{\int_{N-n}^N \frac{q(n')}{n'} dn'}\right).
\end{equation}
Once these functions are calculated, the constants $C_1$, $C_2$, $C_3$ and $C_4$ are obtained by matching $\Pi_\text{in}$ to $\Pi_\text{mid}$ when $1/N \ll x \ll 1$ and $\Pi_\text{out}$ to $\Pi_\text{mid}$  when $1/N \ll 1-x \ll 1$. Clearly, $\sigma_\text{e}$ increases as $x(1-x)$ decreases, since the strength of the drift is added to the environmental fluctuations. Accordingly (see Figure \ref{fig2m}), $q$ at the edges is smaller and its value increases in the bulk.

\begin{figure}[h]
\centering
\includegraphics[width=0.45\textwidth]{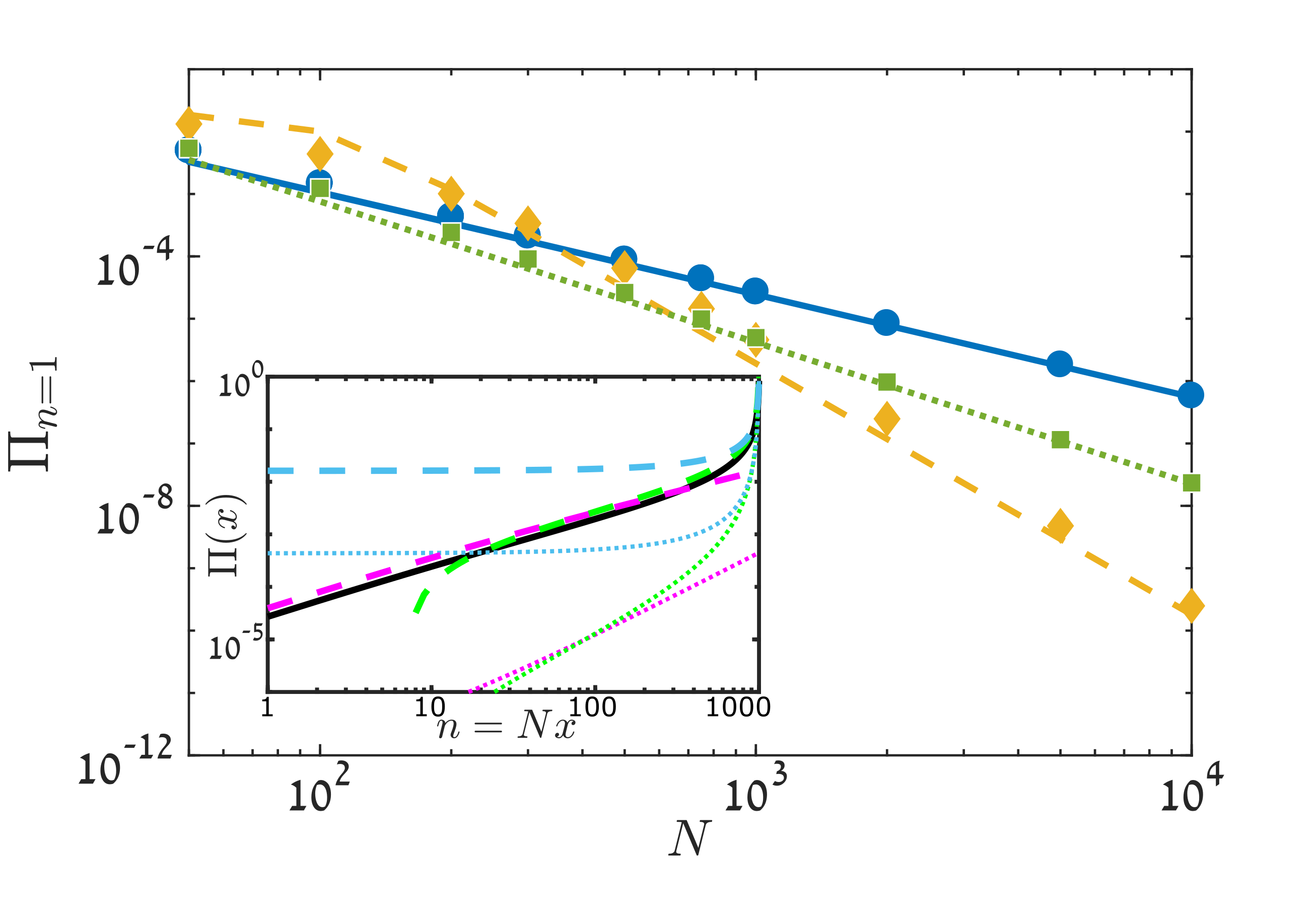}
\caption{\label{single8} The inset shows the chance of fixation as a function of $n$ (double logarithmic scale) for $s_0 = -0.1$ and $\sigma = 0.5$. Inner (magenta), intermediate (green) and outer (blue) solutions are compared with the result of a direct numerical solution of Eq.~(\ref{eqn1}) (solid black line). Clearly, the diffusion approximation (dotted curves) performs poorly while the small-$q$ approximation (dashed curves) works well.  In the main panel the chance of fixation of a single mutant, $\Pi_{n=1}$, is plotted vs.\ $N$. Symbols are the results of direct numerical solution, while lines are the prediction of Eq.~(\ref{single4}). The parameters are $s_0 = -0.1$, $\sigma = 0.5$ (full line, circles); $s_0 = -0.05$, $\sigma = 0.3$ (dotted line, triangles); and $s_0 = -0.01$, $\sigma = 0.1$ (dashed line, diamonds). The theory captures the general trend well, and the relative error decreases with $N$. To improve the quality of the approximation, the parameter $q_\text{mid}$ has been taken from the direct numerical solution of Eq.~(\ref{eqn2a}) with $s_\text{e} = s_0$ and $\sigma_\text{e} = \sigma$. }
\end{figure}

In the simplest scenario $|q_\text{mid}| < 0.25$, so for any $z$ the system is in the small-$q$ sector. In this case the special solution in the inner regime is found by integrating over
\begin{equation}\label{qinmain}
q_\text{in} = -\frac{2 s_0 n}{(s_0^2+\sigma^2)n + e^{-2s_0}\cosh(2\sigma)},
\end{equation}
and the chance of fixation for a single individual is
\begin{align} \label{single4}
\Pi_{n=1} = \frac{\left[1+\frac{e^{2s_0}(s_0^2+\sigma^2)}{\cosh(2\sigma)}\right]^{-\frac{2s_0}{s_0^2+\sigma^2}}-1}
{\left[\frac{N(s_0^2+\sigma^2)}{\cosh(2\sigma)}\right]^{-\frac{4s_0}{s_0^2+\sigma^2}}-1}.
\end{align}
Figure~\ref{single8} shows that Eq.~(\ref{single4}) works well, and the small-$q$ solution is much better than the DA solution of \cite{danino2018fixation,meyer2018noise}.

When $q$ in the intermediate regime is not small ($|s_\text{e}/\sigma_\text{e}| >0.25$) one may guess an inner solution that converges to the appropriate limit when $n \to 0$ and matches the intermediate solution for $n \to \infty$. In the companion paper~\cite{comp} we suggest such an expression that works well in the medium-$q$ sector, and sketch the ways in which one may produce better approximations.

An important aspect of our analysis has to do with the distinction between weak and strong selection.  In a fixed environment, if $|s|N \ll 1$ (weak selection) the dynamics are neutral. If $|s|N \gg 1$ (strong selection) the dynamics are still approximately neutral (drift-dominated) up to $n_\text{c} = 1/(2|s|)$ and is nearly deterministic above this point, so the chance of fixation for a beneficial mutant becomes $N$-independent once $N>n_\text{c}$~\cite{desai2007speed}. For a deleterious mutant $\Pi$ is exponentially small in $N$ as long as $N>n_\text{c}$.

Analogously, in a stochastic environment $n_\text{c}$ is the point at which  $S_n  \approx 1$. If we assume that $n_\text{c}$ happens to lie in the inner regime where $z = \ln x$ (but still $n_\text{c} \gg 1$), then $S = \int q dz = \int (q/n) dn  $. In the inner regime $q$ is usually small, so $S \approx 2s_0 \ln[1+n(s_0^2+\sigma^2)]/(s_0^2 + \sigma^2)$. As a result,
\begin{equation}
n_\text{c} = \frac{e^\frac{s_0^2+\sigma^2}{2s_0}-1}{s_0^2+\sigma^2}.
\end{equation}
This expression generalizes a similar criterion suggested in~\cite{cvijovic2015fate}. The two expressions coincide when DA holds, i.e., when $s_0^2$ is neglected with respect to $\sigma^2$.  The pure drift expression $1/(2|s_0|)$ emerges when $\sigma \to 0$. $n_\text{c}$ diverges exponentially when $|s_0| \to 0$, meaning that even huge populations may be in the weak selection regime. In general, to obtain an expression for $n_\text{c}$ when our assumptions do not hold, one has to calculate $S$ for a given $q$ and to apply the condition $S(n_\text{c}) =1$.

Importantly, in a stochastic environment weak selection does not imply neutrality. Instead, in the weak-selection regime the system is in a ``time-averaged neutral"~\cite{kalyuzhny2015neutral,danino2018theory} phase, meaning that $s_\text{e}$ is effectively zero so both types have the same mean fitness (when  fitness is averaged over time). In this case the abundance undergoes an unbiased random walk along the $z$-axis, so the chance of fixation at $n$ behaves in general like $z/z_\text{max}$ and decays logarithmically with $N$. Only when $\sigma \sqrt{N} \ll 1$ does the drift (demographic noise) dominate the fluctuating selection and Kimura's neutral dynamics (in which the chance of fixation is $n/N$) are restored.

The method presented here relies on three approximations. The leading-order WKB and the asymptotic matching techniques are well-studied procedures, with corrections that decay to zero as $N$ increases. The two-destination approximation is less controlled and breaks down when $\sigma \to |s_0|$. Still, as demonstrated, the range of applicability of our method is much wider than that of DA. Some possible solutions to the technical problems associated with the two-destination procedure are discussed in the companion paper~\cite{comp}.

Putting aside some laboratory setups, temporal environmental stochasticity plays a major role all across Nature and its effect is typically strong, rendering DA ineffectual. We hope that the method presented here will facilitate  the quantitative assessment of parameters like fixation probability, invasion probability and persistence times, which govern phenotypic and species diversity on Earth.

\bibliography{refs_ives}

\end{document}